\documentclass[useAMS,usenatbib]{mnras}
\usepackage{enumitem}
\usepackage{graphicx}
\usepackage{amsmath, amssymb}
\usepackage[T1]{fontenc}
\usepackage[usenames]{color}
\usepackage[dvipsnames]{xcolor}
\usepackage{bm}
\usepackage[capitalize]{cleveref}
\usepackage{physics}
\usepackage{xspace}
\usepackage{acro}
\usepackage{caption}
\usepackage{subcaption}

\newcommand{\Mpeak}{M_\mathrm{peak}}
\newcommand{\Mvir}{M_\mathrm{vir}}
\newcommand{\Vmaxhalo}{V_\mathrm{max,halo}}
\newcommand{\Vmax}{V_\mathrm{max}}
\newcommand{\Vflat}{V_\mathrm{flat}}
\newcommand{\sint}{\sigma_\mathrm{SHAM}}

\newcommand{\Mpch}{h^{-1}\,\mathrm{Mpc}}
\newcommand{\kpch}{h^{-1}\,\mathrm{kpc}}
\newcommand{\Msunph}{h^{-1}\,M_\odot}
\newcommand{\kmsecMpc}{\mathrm{km}\,\mathrm{s}^{-1}\,\mathrm{Mpc}^{-1}}

\defcitealias{Stiskalek_2021}{S21}

\DeclareRobustCommand{\VAN}[3]{#2}
\let\VANthebibliography\thebibliography
\def\thebibliography{\DeclareRobustCommand{\VAN}[3]{##3}\VANthebibliography}

\newcommand{\bibnote}[2]{\global\@namedef{#1note}{#2}}
\newcommand{\biblink}[2]{\global\@namedef{#1link}{#2}}

\makeatletter
\DeclareRobustCommand{\HI}{%
  \mbox{H\check@mathfonts\fontsize\sf@size\z@\selectfont I}%
}
\makeatother

\title[Testing SHAM with galaxy kinematics]{Testing subhalo abundance matching with galaxy kinematics}
\author[]{Fedir Boreiko\thanks{Email: \href{mailto:fedir.boreiko@gmail.com}{fedir.boreiko@gmail.com}}$^{1,2,3}$, Tariq Yasin$^1$, Harry Desmond$^4$, Richard Stiskalek$^1$ and Matt J. Jarvis$^{1,5}$
\\
$^{1}$Astrophysics, University of Oxford, Denys Wilkinson Building, Keble Road, Oxford, OX1 3RH, UK\\
$^{2}$Institute of Astronomy, University of Cambridge, Madingley Road, Cambridge CB3 0HA, UK\\
$^{3}$Department of Physics and Astronomy, University of Manchester, Manchester M13 9PL, UK\\
$^{4}$Institute of Cosmology \& Gravitation, University of Portsmouth, Dennis Sciama Building, Portsmouth, PO1 3FX, UK\\
$^{5}$Department of Physics and Astronomy, University of the Western Cape, Robert Sobukwe Road, 7535 Bellville, Cape Town, South Africa}

\pubyear{\the\year{}}

\begin{document}\label{firstpage}
\pagerange{\pageref{firstpage}--\pageref{lastpage}}
\maketitle

\begin{abstract}
The rotation velocities of disc galaxies trace dark matter halo structure, providing direct constraints on the galaxy--halo connection. We construct a Bayesian forward model to connect the dark matter halo population predicted by $\Lambda$CDM with an observed sample of disc galaxies (SPARC) through their maximum rotation velocities. Our approach combines a subhalo abundance matching scheme (accounting for assembly bias) with a parameterised halo response to galaxy formation. When assuming no correlation between selection in the SPARC survey and halo properties, reproducing the observed velocities requires strong halo expansion, low abundance matching scatter ($<0.15$ dex at $1\sigma$) and a halo proxy that strongly suppresses the stellar masses in satellite haloes. This is in clear tension with independent clustering constraints. Allowing for SPARC-like galaxies to preferentially populate low $\Vmax$ haloes at fixed virial mass greatly improves the goodness-of-fit
and resolves these tensions: the preferred halo response shifts to mild contraction, the abundance matching scatter increases to $\sint = 0.19^{+0.13}_{-0.11}$ dex and the proxy becomes consistent with clustering. However, the inferred selection threshold is extreme, implying that SPARC galaxies occupy the lowest ${\sim}16$ per cent of the $\Vmaxhalo$ distribution at fixed $\Mvir$. Moreover, even with selection, the inferred scatter remains in statistical disagreement with the low-mass clustering constraints, which are most representative of the SPARC galaxies in our sample. Our analysis highlights the advantage of augmenting clustering-based constraints on the galaxy--halo connection with kinematics and suggests a possible tension using current data.
\end{abstract}

\section{Introduction}

In $\Lambda$CDM, every galaxy resides within a dark matter halo whose properties---mass, concentration, formation history---shape the galaxy's observable characteristics. Constraining this ``galaxy--halo connection'' empirically is essential both for building theoretical models of galaxy formation and for using observed galaxy populations to infer the underlying dark matter distribution~\citep{Wechsler2018}. A particularly direct probe comes from galaxy kinematics: the internal motions of stars and gas trace the total gravitational potential, and hence the dark matter content, of individual systems.

Rotation curves---the circular velocity of disc galaxies as a function of galactocentric radius---provide detailed constraints on the radial mass distribution. \HI\ observations revealed that rotation curves remain flat or rising well beyond the optical disc~\citep{Rubin1970, Bosma1981}, implying mass profiles that extend far beyond the visible baryons. This established dark matter haloes as a ubiquitous component of disc galaxies~\citep{SofueRubin2001}. In recent years, the SPARC database~\citep{Lelli_2016} has provided high-quality rotation curves for 175 nearby galaxies, combining \HI\ interferometry with Spitzer $3.6~\mu$m photometry~\citep{Werner2004} to enable precise decomposition into baryonic and dark matter contributions.

Rotation curves are often summarised by characteristic velocity statistics such as the peak velocity $\Vmax$, the asymptotic flat velocity $\Vflat$, or the velocity at a fixed multiple of the disc scale length $V_{2.2}$~\citep{Ponomareva_2017, Ponomareva_2017b, Lelli_2019}. These summary statistics correlate tightly with baryonic mass: the Baryonic Tully--Fisher Relation (BTFR) connects a galaxy's total baryonic mass to its rotational velocity with remarkably small scatter ($\sim$0.1 dex;~\citealt{Tully1977, McGaugh2000, Lelli_2019, Desmond2017}). This tightness places strong constraints on the galaxy--halo connection and has been used extensively to test dark matter models and galaxy formation physics~\citep{TrujilloGomez2011, Desmond_2015, Desmond2019, Ferrero2017}.

Several studies have used subhalo abundance matching (SHAM) to forward-model galaxy scaling relations such as the BTFR~\citep{TrujilloGomez2011, Desmond_2015, Ferrero2017}. SHAM provides a framework for connecting galaxies to haloes in dark-matter-only simulations by rank-ordering both populations---typically by luminosity or stellar mass, and a halo property such as present-day virial mass $\Mvir$ or the peak mass across a halo's history $\Mpeak$---and matching them with some scatter~\citep{Conroy_2006, Behroozi_2010, Moster_2010}. The intrinsic scatter $\sint$ (typically $\sim 0.2\,\mathrm{dex}$;~\citealt{Reddick_2013, Desmond_2015}) is routinely constrained using galaxy clustering. SHAM successfully reproduces galaxy clustering, satellite fractions, and other statistics, providing strong constraints on the stellar-to-halo mass relation~\citep{Reddick_2013}.

Modern abundance matching employs multi-property halo proxies to account for assembly bias---correlations between clustering and secondary halo properties at fixed mass---that can bias inferences if ignored~\citep{Hearin2013,Zentner2014,Mao_2017}. \citet{Lehmann_2017} introduced a proxy that smoothly interpolates between peak halo mass $\Mpeak$ and peak maximum circular velocity $V_\mathrm{max,peak}$, controlling sensitivity to halo concentration at fixed mass. The interpolation is controlled by a single free parameter, which they constrained along with the scatter using clustering data (their best-fit values were subsequently tested against kinematics by \citealt{Desmond2017}). This accommodates assembly bias within the SHAM framework.

In this work, we improve on previous SHAM-based models of galaxy kinematics in two ways. First, rather than using a fixed SHAM proxy, we adopt the compound proxy of~\citet[][hereafter~\citetalias{Stiskalek_2021}]{Stiskalek_2021}, which interpolates between $\Mvir$ and $\Mpeak$, and allow its parameter $\alpha$ to be constrained along with the SHAM scatter in the inference. This affords a test of whether the galaxy--halo connection inferred from clustering is consistent with that inferred from galaxy dynamics. Second, unlike previous works that fit population-level summary statistics such as the BTFR slope, scatter and intercept~\citep{TrujilloGomez2011, Desmond_2015, Ferrero2017,Desmond2017}, we construct a full likelihood for the observed maximum circular velocity $\Vmax$ of \emph{each individual galaxy} in SPARC, maximising the information extracted from the galaxy--halo connection without compression into the summary statistics of the TFR properties.

We use $\Vmax$ as our kinematic observable for two reasons. First, although $\Vflat$ is the quantity that correlates best with baryonic properties~\citep{Lelli_2019}, it is more difficult to measure than $\Vmax$, requiring data at higher galactocentric radius as well as a subjective definition of what constitutes ``flatness''. Second, $\Vmax$ is close in information content to the \HI\ line width $W_{50}$~\citep{Yasin2023b}, a velocity summary statistic that will be available for millions of galaxies at cosmological redshifts from current and upcoming \HI\ surveys such as MIGHTEE~\citep{Jarvis2016,Ponomareva2021, Varasteanu2025}, WALLABY~\citep{Koribalski2020}, LADUMA~\citep{Blyth2016}, DINGO~\citep{Rhee2023} and eventually, the SKA~\citep{Blyth2015}. Developing and validating forward models using $\Vmax$ therefore enables application to much larger samples at higher redshift.

Alongside the SHAM parameters, we consider free parameters governing the response of the dark matter halo to galaxy formation (spanning contraction to expansion) and a ``selection parameter'' allowing SPARC galaxies to occupy a biased subset of the halo population. Specifically, we ask:
\begin{itemize}
    \item What values of the SHAM parameters ($\alpha$, $\sint$) are preferred by SPARC kinematics, and are these consistent with clustering constraints?
    \item Is halo expansion or contraction required to match the observed velocities?
    \item To what extent is selection important for modelling the kinematics, and how does this impact the consistency with inferences from clustering?
    \item What do these results imply for the host haloes of gas-rich, late-type galaxies?
\end{itemize}
The paper is structured as follows. Section~\ref{sec:data} describes the observational and simulation data we use. Section~\ref{sec:methods} details our forward-modelling methodology: SHAM, rotation curve calculation including halo response, selection modelling and the likelihood framework. Section~\ref{sec:results} presents results from mock validation and fits to the real data. We discuss implications and compare to the literature in Section~\ref{sec:discussion}, and conclude in Section~\ref{sec:conclusion}.
We denote base-10 logarithms as $\log$ and natural logarithms as $\ln$.

\section{Observed and Simulated Data}\label{sec:data}

\subsection{SPARC}\label{sec:sparc}

Our kinematic data are derived from the SPARC database\footnote{\url{https://astroweb.case.edu/SPARC/}}~\citep{Lelli_2016}. This contains 175 late-type galaxies for which both spatially-resolved \HI\ and H$\alpha$ rotation curves and near-infrared Spitzer photometry are available, providing a detailed view of the distribution of stars and gas.

The sample covers a wide range of luminosities (from $10^7$ to $10^{12}\,L_\odot$), surface brightness ($\sim$5 to 5000 $L_\odot\,\mathrm{pc}^{-2}$), gas mass (from $10^7$ to $10^{10.6}\,\mathrm{M}_\odot$), and morphology (from S0 to Im/BCD types). At $3.6~\mu$m, the mass-to-light ratio is assumed to remain relatively stable~\citep{McGaugh_Schombert_2014}, aiding in disentangling the velocity contributions of stellar mass and dark matter (but see~\citealt{Varasteanu2025} for evidence that this may not be universally true).

Within our likelihood framework, we compare predicted $\Vmax$ values with those measured from SPARC rotation curves to infer model parameters. This comparison is based on a sub-sample of 153 galaxies used in SPARC BTFR analyses, where galaxies with low inclinations and low-quality rotation curves have been excluded~\citep{Lelli_2019}.
We adopt the SPARC observational uncertainties on $\Vmax$, $\sigma_{\Vmax}^{\text{(SPARC)}}$, which arise from rotation curve measurements and disc inclination; the latter typically dominates~\citep{Lelli_2019}. We treat these as Gaussian and uncorrelated between galaxies.

\subsection{NASA-Sloan Atlas}\label{sec:nsa}

We obtain the observed stellar mass function (SMF) from the NASA-Sloan Atlas\footnote{\url{https://www.sdss4.org/dr17/manga/manga-target-selection/nsa/}} (NSA), a catalogue of nearby galaxies combining the Sloan Digital Sky Survey (SDSS) with the Galaxy Evolution Explorer~\citep{Martin_2005}. We use \texttt{NSA v1\_0\_1}, based on SDSS DR13~\citep{Albareti_2017}, which contains $\sim$640,000 galaxies out to redshift $z = 0.15$. The catalogue includes both elliptical Petrosian and S\'ersic aperture photometry, K-corrected to $z = 0.0$, with the former considered more reliable.
We therefore adopt the elliptical Petrosian SMF following~\citetalias{Stiskalek_2021}, computed using the $1/V_\mathrm{max,vol}$ method~\citep{Schmidt_1968}. This corrects for Malmquist bias by weighting each galaxy by the inverse of the maximum comoving volume within which it is detectable.

Our abundance matching uses the NSA SMF, yet only 91 SPARC galaxies have NSA stellar masses. For these, the NSA $M_\star$ estimates are statistically consistent with SPARC values assuming ${\sim}0.2$ dex uncertainties on NSA values~\citep{Yasin2023a}. This consistency justifies using SPARC stellar masses for individual galaxies while employing the NSA-derived SMF for the input to SHAM.

\subsection{Simulation Data}\label{sec:simdata}

We use the $140~\Mpch$ ``Shin-Uchuu'' box from the Uchuu suite of cosmological $N$-body simulations~\citep{Ishiyama_2021}, run with the GreeM $N$-body code~\citep{Ishiyama_2009, Ishiyama_2012}. The simulation includes $6400^3$ particles, each with a mass of $8.97 \times 10^5~\Msunph$ and a force softening length of $0.4~\kpch$, and adopts the 2018 \textit{Planck} flat $\Lambda$CDM cosmology~\citep{Planck_2020}: $H_0 = 67.74~\kmsecMpc$, $\Omega_\mathrm{m} = 0.3089$, $\Omega_\Lambda = 0.6911$, $n_s = 0.9667$, and $\sigma_8 = 0.8159$.
Haloes and satellite haloes are identified with the \texttt{Rockstar} halo finder~\citep{Behroozi_2013}, with merger trees constructed using \texttt{Consistent Trees}~\citep{Behroozi_2013_b}. Halo masses are defined by the virial overdensity $\Delta_{\text{vir}} = 178$~\citep{Bryan_1998}. Our sample includes haloes with $\Mvir > 10^{10}\,\mathrm{M}_\odot$.

\section{Methods}\label{sec:methods}

We summarise our method for modelling galaxy rotation curves below; each step is detailed in the subsequent sections.
\begin{enumerate}
    [label=\hspace{\parindent}(\roman*), align=left, leftmargin=0pt, labelsep=0.5em]
    \item Apply SHAM to assign stellar masses to haloes using a generalised proxy parametrised by $\alpha$ and intrinsic scatter $\sint$ (Section~\ref{sec:sham}).
    \item Optionally, select haloes by their maximum circular velocity $\Vmaxhalo$ to model how selection effects influence $\alpha$, since SPARC may be a biased sample of the halo population (Section~\ref{sec:selection}).
    \item Generate Monte Carlo realisations of galaxy stellar masses by sampling their observational uncertainties, and associate them with haloes via SHAM.
    (Section~\ref{sec:dynamics}).
    \item Modify halo density profiles to account for disc formation---ranging from adiabatic contraction to expansion---then compute rotation curves and extract $\Vmax$ (Section~\ref{sec:halo_response}).
    \item Repeat the above steps by resampling the SHAM galaxy--halo connection varying the noise realisation at fixed $\sint$ to capture stochasticity in the stellar--halo mass relation, collecting $\Vmax$ samples across realisations.
    \item Compare each galaxy's $\Vmax$ distribution with observations to evaluate the likelihood; the total likelihood is then the product over all galaxies (Section~\ref{sec:likelihood}).
    \item Evaluate the likelihood on a grid of model parameters.
    \item Interpolate the likelihood grid and sample posterior distributions of the model parameters using Markov Chain Monte Carlo (MCMC).
\end{enumerate}

\subsection{Modelling \texorpdfstring{$\Vmax$}{Vmax}}

\subsubsection{Subhalo abundance matching}\label{sec:sham}

SHAM is an empirical framework that associates galaxies with dark matter haloes by assuming a monotonic relation between (e.g.) stellar mass and a halo property (the ``halo proxy''). Haloes are rank-ordered by the chosen proxy, while galaxies are rank-ordered by stellar mass using an observed SMF (here from the NSA). Galaxies and haloes are then matched by rank: the galaxy with the $n$\textsuperscript{th} largest stellar mass is assigned to the halo with the $n$\textsuperscript{th} largest proxy value~\citep{Kravtsov_2004, Vale_Ostriker_2004, Conroy_2006, Behroozi_2010, Moster_2010}. SHAM reproduces observed galaxy clustering, galaxy--galaxy lensing, group catalogues, and void statistics~\citep{Conroy_2006, Vale_Ostriker_2006, Moster_2010, Reddick_2013}, as well as stellar--halo mass relations inferred from weak lensing and satellite kinematics~\citep{Behroozi_2010}.

SHAM can incorporate non-zero scatter $\sint$, the standard deviation of stellar mass at fixed halo proxy~\citep{Tasitsiomi_2004, Behroozi_2010}. This scatter largely reflects the influence of secondary halo properties on baryonic content, particularly accretion history~\citep{Tinker_2017}. We treat $\sint$ as a free parameter and implement scatter using the deconvolution method of~\citet{Behroozi_2010}: galaxies are first matched at zero scatter, stellar masses are perturbed along the galaxy--halo ladder, and abundances are adjusted to restore consistency with the input SMF at chosen $\sint$.

We adopt the generalised halo proxy $m_{\alpha}$ from~\citetalias{Stiskalek_2021}, who found it to perform significantly better than the~\citet{Lehmann_2017} proxy for faint optically selected samples, while performing equally well for bright samples. This proxy interpolates between peak virial mass $\Mpeak$ and present-day virial mass $\Mvir$,
\begin{equation}
    m_{\alpha} = \Mvir \biggl(\frac{\Mpeak}{\Mvir} \biggr)^{\alpha},
\end{equation}
where $\alpha$ is a free parameter. For $\alpha = 0$, haloes are ranked by $\Mvir$ and for $\alpha = 1$ by $\Mpeak$, with larger $\alpha$ boosting the clustering signal. Varying $\alpha$ controls the ranking of haloes that have lost mass (typically satellites): $\alpha > 0$ up-ranks 
haloes whose mass peaked in the past, while $\alpha < 0$ down-ranks them.

\subsubsection{Forward modelling the BTFR}\label{sec:dynamics}

We generate $N_{\text{stellar}}$ Monte Carlo samples for each SPARC galaxy to estimate stellar and gas masses from the photometric data. First, we draw $N_{\text{stellar}}$ samples of distance, $L_{[3.6]}$, and $M_{\text{HI}}$ from Gaussian distributions centred on the SPARC values with standard deviations set by measurement uncertainties. Adopting mass-to-light ratios $\Upsilon_\star = 0.5\,M_\odot/L_\odot$ (disc) and $\Upsilon_\star = 0.7\,M_\odot/L_\odot$ (bulge), each with 0.1~dex uncertainty~\citep{Lelli_2016}, we sample $\Upsilon_\star$ similarly. Stellar and gas masses are computed as
\begin{equation}
\begin{aligned}
    &M_\star = (L_{[3.6]} - L_{\text{bulge}}) \Upsilon_{\rm disc}
    + L_{\text{bulge}} \Upsilon_{\rm bulge}, \\
    &M_{\text{gas}} = 1.33\,M_{\text{HI}},
\end{aligned}
\end{equation}
where $L_{\text{bulge}}$ is the bulge luminosity from photometric decomposition\footnote{More realistically, $L_{\mathrm{bulge}}$ should scale with $L_{[3.6]}$, but as the luminosity uncertainties are small compared to other uncertainties, especially the mass-to-light ratios, this does not make an appreciable difference.}. We rescale masses to the sampled distance, as SPARC tabulates luminosities/masses assuming the fiducial distances to each galaxy. This yields $N_{\text{stellar}}$ samples of $M_\star$ and $M_{\text{gas}}$ for each galaxy.

We assign each $M_\star$ sample to a halo by matching to the nearest stellar mass in the SHAM catalogue, yielding $N_{\text{stellar}}$ galaxy--halo pairs per galaxy. This propagates $M_\star$ uncertainties into the halo assignment and incorporates SHAM scatter, so that the pairing reflects both observational uncertainties and intrinsic scatter in the galaxy--halo connection.

The total circular velocity is the quadrature sum of contributions from dark matter, stellar bulge, stellar disc, and gas disc,
\begin{equation}
\begin{aligned}
V_C^2 &= V_{\text{dm}} \lvert V_{\text{dm}} \rvert
      + V_{\text{disc}} \lvert V_{\text{disc}} \rvert \\
     &\quad + V_{\text{bulge}} \lvert V_{\text{bulge}} \rvert
      + V_{\text{gas}} \lvert V_{\text{gas}} \rvert ,
\end{aligned}
\end{equation}
where $V_{\text{disc}}$ and $V_{\text{bulge}}$ are scaled from the fiducial SPARC stellar velocity contributions (computed assuming $\Upsilon = 1\,M_\odot/L_\odot$~\citep{Lelli_2016}) according to the sampled mass-to-light ratios:
\begin{equation}
\begin{aligned}
    V_{\text{disc}} &= \sqrt{\Upsilon_{\text{disc}}} \, V_{\text{disc, fid}} \\
    V_{\text{bulge}} &= \sqrt{\Upsilon_{\text{bulge}}} \, V_{\text{bulge, fid}}
\end{aligned}
\end{equation}

For the dark matter contribution, we assume the Navarro--Frenk--White profile (NFW;~\citealt{Navarro_1997})
\begin{equation}
    \rho_{\text{NFW}}(r) = \frac{\rho_s}{(r/r_s)[1 + (r/r_s)]^2},
\end{equation}
where $r_s$ is the scale radius and $\rho_s$ the characteristic density. The enclosed mass within radius $r$ is
\begin{equation}
    M_{\text{NFW}}(r) = 4\pi\rho_s r_s^3 \left[\ln(1 + x) - \frac{x}{1 + x}\right],
\end{equation}
where $x = r/r_s$. The circular velocity contribution from dark matter is
\begin{equation}
    V_{\text{dm}}(r) = \sqrt{\frac{GM_{\text{NFW}}(r)}{r}}.
\end{equation}
From each rotation curve we extract $\Vmax$, defined as the peak circular velocity, and compute $M_{\text{bar}} = M_\star + M_{\text{gas}}$. For a given SHAM realisation, this yields $N_{\text{stellar}}$ samples of $(\Vmax, M_{\text{bar}})$ per galaxy. We repeat the procedure for $N_{\text{SHAM}}$ SHAM realisations at fixed $(\alpha, \sint)$, resampling the noise each time to capture stochasticity in the galaxy--halo connection.

\begin{figure}
    \centering
    \includegraphics[width=\columnwidth]{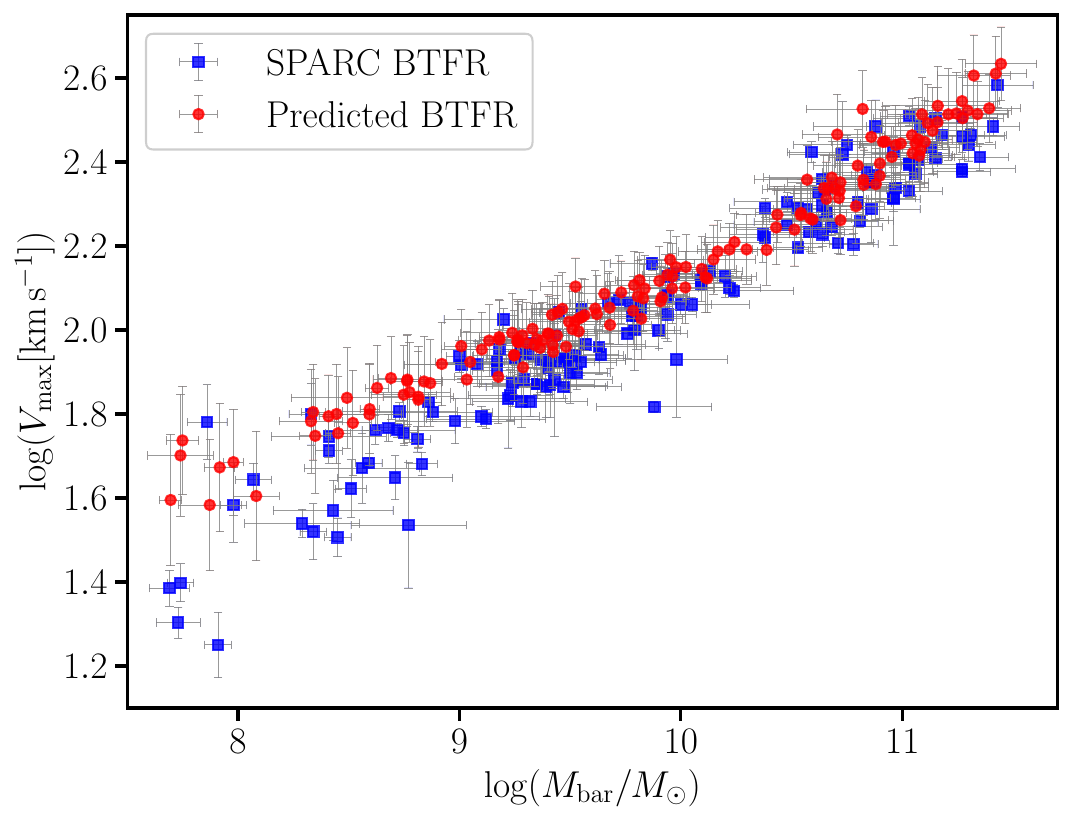}
    \caption{Comparison of forward-modelled (red) and observed SPARC (blue) BTFRs. Model parameters are $\tan^{-1}\alpha = 0.5$, $\sint = 0.1\,\mathrm{dex}$, $\nu = -1.1$, with no halo selection ($x = 0$).
    }
    \label{fig:btfr_example}
\end{figure}

Collecting all $N_{\text{stellar}} \times N_{\text{SHAM}}$ samples per galaxy, we transform $\Vmax$ and $M_{\text{bar}}$ to $\log$.
\Cref{fig:btfr_example} shows the predicted BTFR for illustrative parameters and compares it to the SPARC data.

\subsubsection{Halo response}\label{sec:halo_response}

Dark matter haloes in $N$-body simulations agree well with the NFW profile, but galaxy evolution modifies this structure. The classic expectation is adiabatic contraction: as baryons cool and collapse into a central disc, they deepen the potential and draw in surrounding dark matter, steepening the inner profile~\citep{Blumenthal_1986, Gnedin_2004, Gnedin_2011}. However, bursty star formation and feedback-driven outflows can transfer energy to the dark matter and cause the opposite effect---halo expansion and core formation~\citep{Pontzen_Governato_2012}. The net response depends on the balance between these processes, which varies with halo mass and star formation efficiency: cores form most readily at intermediate stellar-to-halo mass ratios where feedback is energetic relative to the binding energy, while contraction dominates in more massive systems where baryons deepen the potential faster than feedback can unbind material~\citep{DiCintio2014, Tollet2016}. Recent work calibrating quasi-adiabatic relaxation models against IllustrisTNG~\citep{Pillepich2018} and EAGLE~\citep{Schaye2015} suggests that the response may additionally depend on halo-centric distance~\citep{Velmani2022}, though this is sensitive to the subgrid physics adopted in those simulations.

Given these uncertainties, we adopt a simple parameterisation that spans the full range from contraction to expansion. We use the modified adiabatic contraction framework of~\citet{Gnedin_2011}, which accounts for the eccentricity of dark matter particle orbits, but introduce a parameter $\nu$ that continuously interpolates between contraction ($\nu > 0$) and expansion ($\nu < 0$)~\citep{Dutton_2007, Desmond_2015}.

In this model, we assume that before disc formation, both dark matter and baryons follow the NFW density profile exactly, with baryons initially tracing the dark matter distribution. Upon disc formation, baryons are redistributed into a thin exponential disc. Assuming the halo is spherically symmetric and composed of non-crossing shells, angular momentum is conserved,
\begin{equation} \label{eq:MAC}
    \left[M_{\text{dm},i}\left(\overline r_i\right) + M_{\text{b},i}\left(\overline r_i\right)\right]r_i =
    \left[M_{\text{dm},i}\left(\overline r_i\right) + M_{\text{b},f}\left(\overline r_f\right)\right]r_f,
\end{equation}
where subscripts $i$ and $f$ refer to the states before and after disc formation, respectively, and the masses are enclosed within the orbit-averaged radius $\overline{r}$. Here we use conservation of dark matter mass within a Lagrangian shell, $M_{\text{dm},f}\left(\overline r_f\right)=M_{\text{dm},i}\left(\overline r_i\right)$. The equation for $\overline r$ is approximated by a power law
\begin{equation}
    \overline r = A_0r_0\left(\frac{r}{r_0}\right)^w,
\end{equation}
with $A_0 = 1.6$, $w=0.8$, and $r_0=0.03r_{\text{vir}}$~\citep[eq.~4]{Gnedin_2011}, where we take the virial radius of the system to equal the halo virial radius, $r_{\rm vir}$.

Using the baryon fraction $f_\text{b} = M_\text{b} / (M_\text{vir} + M_\text{b})$, we rewrite the initial total enclosed mass as $M_{\text{tot}, i}(r) = M_{\text{dm}, i}(r)/(1-f_\text{b})$. Expressing radii in units of $r_\text{vir}$ as $y\equiv r/r_\text{vir}$, we introduce the baryonic mass fraction,
\begin{equation} \label{eq:Bar mass density}
    m_\text{b}(y) = \frac{M_\text{b}(r)}{M_\text{tot}(r_\text{vir})} = f_\text{b} \frac{M_\text{b}(y)}{M_\text{b}(1)},
\end{equation}
and the dark matter mass fraction,
\begin{equation} \label{eq:DM mass density}
    m_\text{dm}(y) = \frac{M_\text{dm}(r)}{M_\text{tot}(r_\text{vir})} = (1 - f_\text{b}) \frac{M_\text{dm}(y)}{M_\text{dm}(1)}.
\end{equation}
For an initial NFW dark matter distribution, 
\begin{equation}
    M_{\text{dm}}(y) = \ln(1+cy) - \frac{cy}{1 + cy},
\end{equation}
where $c$ is the halo concentration. For baryons redistributed into an exponential disc,
\begin{equation}
    M_\text{b}(y) = 1-\left(1+\frac{y}{y_\text{b}}\right)\exp\left(-\frac{y}{y_\text{b}}\right),
\end{equation}
where $y_\text{b} = r_\text{b}/r_\text{vir}$ is the baryon scalelength in units of $r_\text{vir}$.
For simplicity, we assume all baryons follow this profile after disc formation, setting $r_\text{b}=R_\text{eff} / 1.67835$, where $R_\text{eff}$ is the observed effective radius. A more rigorous treatment would separate the disc component and define $r_\text{b}$ as its scalelength, requiring an additional model parameter. Finally, using Equations~\eqref{eq:Bar mass density} and~\eqref{eq:DM mass density}, we rewrite Equation~\eqref{eq:MAC} as
\begin{equation}
    \left[m_{\text{dm},i}(\overline{y}_i) + m_{\text{b},f}(\overline{y}_f)\right] y_f - \frac{m_{\text{dm}, i}(\overline{y}_i)}{1-f_\text{b}} y_i = 0,
\end{equation}
which can be solved iteratively for $r_f$ given halo mass, concentration, virial radius, baryon fraction, and scalelength.

This yields the standard adiabatic contraction solution. To allow for expansion as well as contraction, we adopt the generalised form of~\citet{Dutton_2007}, defining
\begin{equation}
    \Gamma(r_\text{i}) \equiv \frac{r_\text{f}}{r_\text{i}},
\end{equation}
and setting the true final radius as
\begin{equation}
    r_{\text{f}, \text{true}} = \Gamma^{\nu} r_\text{i}.
\end{equation}
The free parameter $\nu$ interpolates between standard adiabatic contraction ($\nu = 1$), no halo response ($\nu = 0$), and expansion by the same factor as standard contraction ($\nu = -1$).
Given an observed radius $r_{\text{f,true}}$, we iteratively solve for the initial radius $r_\text{i}$ and compute the enclosed dark matter mass $M_{\text{dm}}(r_\text{i})$ to evaluate $V_{\text{dm}}$. To reduce computational cost, we precompute a grid over $(c, f_{\text{b}}, r_{\text{b}}, r_{\text{f,true}})$ for each value of $\nu$, storing $M_{\text{dm}}(r_{\text{f,true}})$. During the analysis, we perform multivariate linear interpolation over these grids using \texttt{jax.scipy.ndimage} to obtain $V_{\text{dm}}$.

To quantify the accuracy of this interpolation scheme, we validated the precomputed grids against exact calculations. For 20 values of $\nu$ spanning the prior range $[-3, 3]$, we generated 1000 random test points within the parameter space $(c, f_{\mathrm{b}}, r_{\mathrm{b}}, r_{\mathrm{f}})$ and solved the halo response equations iteratively to determine the true enclosed dark matter mass. Across the majority of the parameter space ($\nu > -2$, $\log c \in [1.0, \, 3.0]$, $\log f_{\mathrm{b}} \in [-2.5, \, -1.0]$, $\log  r_{\mathrm{b}} \in [-2.5, \, -1.5]$, $\log r_{\mathrm{f}} \in [-4, \, 0]$), the interpolation is exceptionally accurate, with a mean percentage error on the enclosed dark matter mass fraction $m_{\mathrm{dm}}(r_{\mathrm{f}})$ of $\lesssim 0.5$ per cent. Non-negligible deviations (${\sim}1$--$2$ per cent) only occur at the lower limit of halo expansion ($\nu \rightarrow -3.0$). We attribute this slight degradation to the complex behaviour of the density profile under strong feedback, where the mapping between initial and final radii becomes highly non-linear. However, given that these errors remain small and that the extreme $\nu$ regime is disfavoured by the data, numerical uncertainties in our posteriors associated with the interpolation are negligible.

\subsubsection{Selection model}\label{sec:selection}

Since the SPARC sample consists of late-type galaxies, these may occupy a biased subset of haloes with distinct structural and formation characteristics compared to the overall halo population. To account for this, we adopt a simple phenomenological halo selection method based on the present-day maximum circular velocity, $\Vmaxhalo$. The choice of $\Vmaxhalo$ as the selection criterion is motivated by its strong correlation with our observable, the galaxy rotation curve maximum. Since $\Vmaxhalo$ encapsulates both halo mass and concentration, it correlates more strongly with galaxy stellar mass than virial mass alone. Furthermore, haloes with high $\Vmaxhalo$ at fixed virial mass tend to have distinct formation histories, such as earlier formation t    imes or higher concentrations, which correlate with enhanced clustering, i.e.\ assembly bias~\citep{Lehmann_2017}.

\begin{figure}
    \centering
    \includegraphics[trim=0 0 0 0, clip, width=\columnwidth]{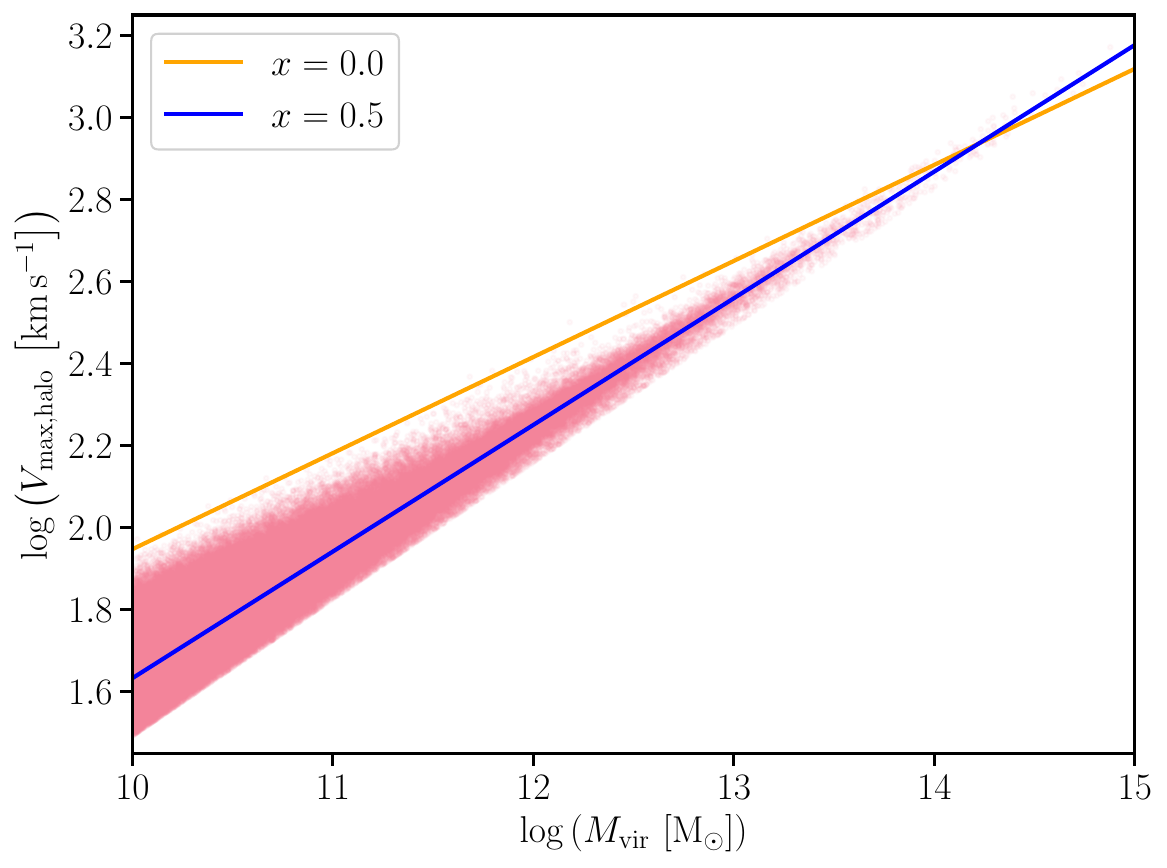}
    \caption{Illustration of the halo-level selection on $\Vmaxhalo$ at fixed $\Mvir$. The solid lines mark the $x = 0$ (\textit{orange}) and $x = 0.5$ (\textit{blue}) thresholds, corresponding to the median $\Vmaxhalo$ at each virial mass. Haloes above this line (higher $\Vmaxhalo$ at fixed mass) would be excluded.}
    \label{fig:halo_selection_plot}
\end{figure}

We consider the halo distribution in $\log\Vmaxhalo$--$\log\Mvir$ space by defining a moving window along the $\log\Mvir$ axis with a width of 0.1 dex. Within each window, we determine the threshold $\log\Vmaxhalo$ value above which a fraction $x$ of the haloes reside. By sliding this window in steps of 0.01 dex along the $\log\Mvir$ axis, we obtain the $x$-percentile of $\log\Vmaxhalo$ as a function of $\log \Mvir$. We then fit a linear model to these percentile points, yielding a continuous linear threshold. We select haloes by removing all that lie above this threshold for a given $x \in [0,1)$. \Cref{fig:halo_selection_plot} illustrates this selection for $x = 0.5$, and \Cref{fig:velocity_distribution_example} shows its effect on the predicted $\Vmax$ distribution.

\subsection{Likelihood framework}\label{sec:likelihood}

We obtain the posterior distribution of the model parameters via Bayes' theorem,
\begin{equation} \label{eq:Bayes}
    \mathcal{P}(\bm{\theta}\mid\mathcal{D}) = \frac{\mathcal{L}(\mathcal{D}\mid\bm{\theta})\mathcal{\pi}(\bm{\theta})}{\mathcal{Z}},
\end{equation}
where $\mathcal{D}$ denotes the data, $\bm{\theta}$ the model parameters, $\mathcal{L}$ the likelihood, $\mathcal{\pi}$ the prior, and $\mathcal{Z}$ the Bayesian evidence,
\begin{equation}
    \mathcal{Z} \equiv \int \mathcal{L}(\mathcal{D}\mid\bm{\theta})\mathcal{\pi}(\bm{\theta}) \, \mathrm{d}\bm{\theta},
\end{equation}
i.e.\ the probability of $\mathcal{D}$ under the chosen model. The Bayes factor (evidence ratio) gives the relative probability of two models given the same data, with qualitative strength often reported on the Jeffreys scale~\citep{jeffreysTheoryProbability1939}.

\begin{figure}
    \centering
    \includegraphics[trim=0 0 0 0, clip, width=\columnwidth]{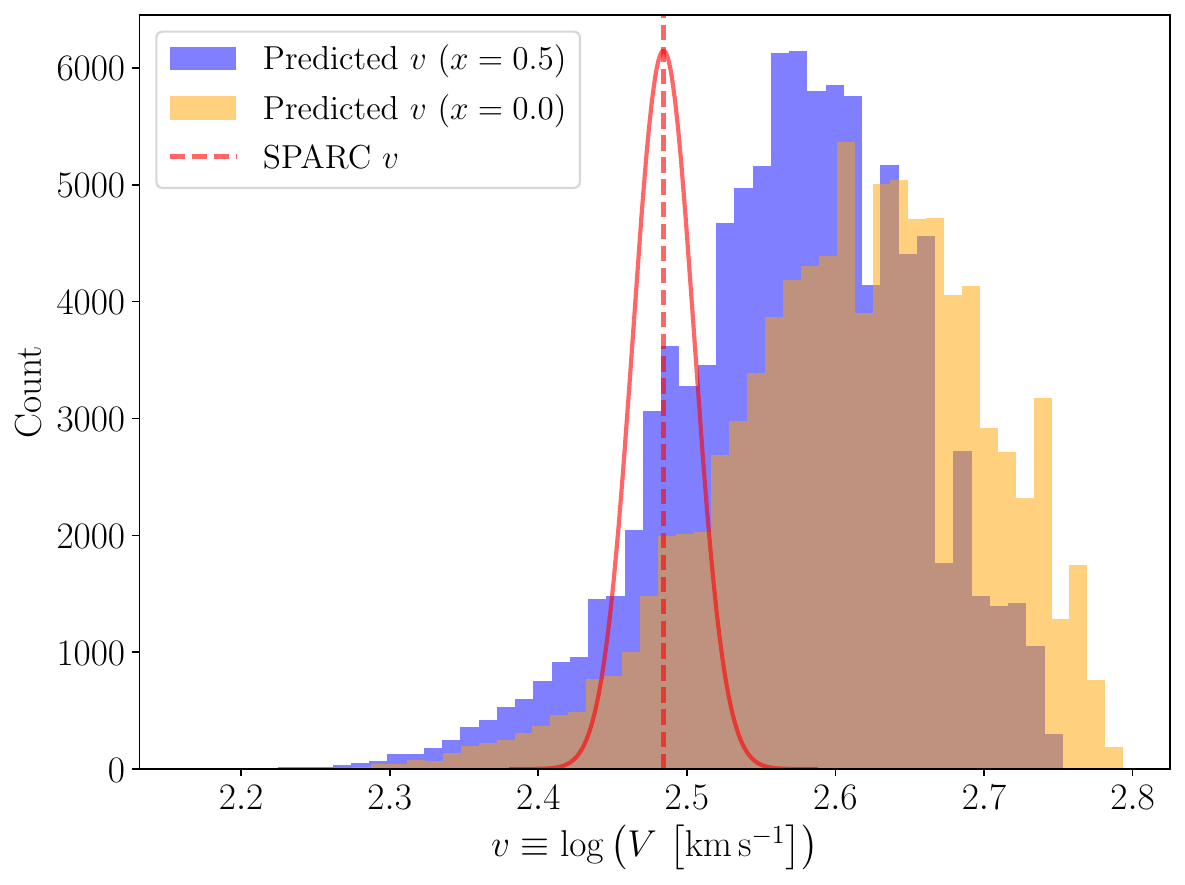}
    \caption{Illustration of the likelihood framework. Distribution of predicted $v \equiv \log \Vmax$ for galaxy UGC02487 in the baseline model without halo selection ($\tan^{-1}\alpha = 0.0$, $\sint=0.1$, $x=0.0$, $\nu=-1.1$) in orange, and with added selection ($\tan^{-1}\alpha = 0.0$, $\sint=0.1$, $x=0.5$, $\nu=-1.1$) in blue, compared to its observed velocity and uncertainty (red). For this example, added selection results in a log-likelihood difference of $\Delta \ln \mathcal{L}(\mathcal{D}_i) \equiv \ln \mathcal{L}(\mathcal{D}_i \mid x=0.5) - \ln \mathcal{L}(\mathcal{D}_i \mid x=0)=0.49$.}
    \label{fig:velocity_distribution_example}
\end{figure}

We perform inference on individual $\Vmax$ values, denoting $v = \log\Vmax$. For the $i$\textsuperscript{th} galaxy, we have the measured SPARC log-velocity $\mathcal{D}_i$ with uncertainty $\sigma_i$, and $N$ predicted values $\{v^{(j)}_i\}_{j=1}^N$.
For a single predicted value $v^{(j)}_i$, the conditional likelihood of observing $\mathcal{D}_i$ is given by a Gaussian likelihood,
\begin{equation}
    \mathcal{L}(\mathcal{D}_i \mid v_i^{(j)},\, \bm{\theta}) = \frac{1}{\sqrt{2\pi}\sigma_i}\exp \left[ -\frac{1}{2} \left( \frac{\mathcal{D}_i - v^{(j)}_i}{\sigma_i} \right)^2\right].
\end{equation}
In the case of a continuous distribution of $v^{(j)}_i$, the marginal likelihood would be
\begin{equation}
    \mathcal{L}(\mathcal{D}_i \mid \bm{\theta}) = \int \mathcal{L}(\mathcal{D}_i \mid v_i,\, \bm{\theta}) p(v_i) \, \mathrm{d}v_i,
\end{equation}
where $p(v_i) \equiv p(v_i \mid \bm{\theta})$ is the distribution of the velocity conditioned on the model parameters, implicitly defined by the forward model. 
We treat it instead as a sum over the Monte Carlo samples:
\begin{equation}
    \mathcal{L}(\mathcal{D}_i\mid\bm{\theta}) \approx \frac{1}{N} \sum_{j=1}^N \mathcal{L}(\mathcal{D}_i \mid v^{(j)}_i,\, \bm{\theta}).
\end{equation}
\Cref{fig:velocity_distribution_example} illustrates this for an example galaxy.
The overall log-likelihood is the sum of the log marginal likelihoods over all galaxies,
\begin{equation}
    \ln\mathcal{L}(\mathcal{D}\mid\bm{\theta}) = \sum_{i=1}^n \ln \mathcal{L}(\mathcal{D}_i\mid\bm{\theta}).
\end{equation}
We evaluate the log-likelihood on a regular grid over $\bm{\theta} = \{\tan^{-1}\alpha, \, \sint, \, \nu, \,x\}$, where we re-parametrise $\alpha$ via $\tan^{-1}\alpha$ to place it on a bounded interval.
We note that this is an example of simulation-based inference (see e.g. \citealt{Cranmer_2020}) as the likelihood function is not known analytically but rather built up from evaluations of the forward model (simulator).

We adopt separable, uniform priors on each parameter (Table~\ref{tab:parameters}):
\begin{itemize}[align=parleft, left=0pt, labelsep=0.5em]
    \item $\tan^{-1}\alpha \sim \text{Uniform}(-\pi/2, \, \pi/2)$. This induces a Cauchy prior on $\alpha$, $p(\alpha) \propto (1+\alpha^2)^{-1}$, which is weakly informative in the tails while avoiding an arbitrary hard truncation in $\alpha$-space. In practice we use a half-open interval $(-\pi/2 + \epsilon, \, \pi/2 - \epsilon)$ with a small $\epsilon$ that prevents the endpoints from mapping to $|\alpha| \rightarrow \infty$ due to floating-point precision.
    \item $\sint \sim \text{Uniform}(0.0, \, 1.0)$ dex. The upper limit is intentionally conservative relative to the ${\sim}0.1$--$0.2$ dex values typically inferred in the literature, ensuring the prior is not informative over the physically relevant range.
    \item $\nu \sim \text{Uniform}(-3.0, \, 3.0)$. This covers a very broad range of halo response, from strong expansion to strong adiabatic contraction. Values $|\nu| \gtrsim 3$ often lead to unphysical or numerically unstable profiles and are disfavoured by existing constraints.
    \item $x \sim \text{Uniform}(0.01, \, 0.99)$. The lower limit avoids the degenerate case of $x = 0$ in our implementation, while the upper limit permits aggressive selection.
\end{itemize}

\begin{table*}
    \centering
    \begin{tabular}{l l p{8.5cm} l}
        \hline
        Parameter & Description & Physical meaning & Prior \\
        \hline
        $\alpha$ & SHAM proxy & Exponent in the generalised halo proxy $m_\alpha {\equiv} \Mvir (\Mpeak/\Mvir)^\alpha$ (see \ref{sec:sham}).  & $\tan^{-1}\alpha \sim \mathrm{Uniform}(-\pi/2,\,\pi/2)$ \\[6pt]
        $\sint$ & Intrinsic scatter & Log-normal scatter in $M_\star$ at fixed halo proxy $m_\alpha$, arising from stochastic galaxy formation processes. & $\mathrm{Uniform}(0.0,\,1.0)$ dex \\[6pt]
        $\nu$ & Halo response & Parameterises dark matter halo response to baryonic infall. $\nu = 0$: no change from NFW. $\nu > 0$: adiabatic contraction ($\nu = 1$ is fiducial). $\nu < 0$: halo expansion (e.g.\ from feedback-driven outflows). & $\mathrm{Uniform}(-3.0,\,3.0)$ \\[6pt]
        $x$ & Selection threshold & Galaxies are drawn only from haloes with $\Vmaxhalo$ below the $x$\textsuperscript{th} percentile of the $\Vmaxhalo$ distribution at fixed $\Mvir$. Higher $x$ implies stronger selection toward low-concentration haloes. & $\mathrm{Uniform}(0.01,\,0.99)$ \\
        \hline
    \end{tabular}
    \caption{Model parameters, physical interpretation, and priors.}
    \label{tab:parameters}
\end{table*}

Between grid points, we interpolate the log-likelihood using multilinear interpolation. From Equation~\eqref{eq:Bayes}, the log-posterior is
\begin{equation}
    \ln \mathcal{P}(\bm{\theta}\mid\mathcal{D}) = \ln\mathcal{L}(\mathcal{D}\mid\bm{\theta}) + \ln\pi(\bm{\theta}) + \text{const.},
\end{equation}
which we sample using an affine-invariant ensemble sampler algorithm implemented in \texttt{emcee} \citep{Foreman_Mackey_2013}. Because the likelihood is computed via Monte Carlo averages, we maintain statistical consistency across $\bm{\theta}$ by ensuring the number of accepted draws per galaxy remains constant on average. Specifically, as the selection threshold $x$ increases and the halo acceptance rate decreases, we scale $N_\text{stellar}$ proportionally to keep the post-selection sample count stable.

To compare consistency of $(\alpha, \sint)$ abundance matching parameters with the independent clustering constraints of~\citetalias{Stiskalek_2021}, we compute the Bayes factor comparing the joint model (in which both datasets constrain the same underlying parameters) to independent models
\begin{equation} \label{eq:Kjoint}
    K_{\mathrm{joint}} = \frac{\mathcal{Z}_\mathrm{joint}}{\mathcal{Z}_\mathrm{SPARC} \, \mathcal{Z}_\mathrm{clustering}}.
\end{equation}
This is also interpreted on the Jeffreys scale, with values of $\log(K_{\mathrm{joint}}) > 0$ indicating consistency and $\log(K_{\mathrm{joint}}) < 0$ tension. Assuming flat priors, this simplifies to a function of the two independent posteriors $p(\bm{\theta})$ and the prior volume ($V$) with $K_{\mathrm{joint}} = V \int p_\mathrm{SPARC}(\bm{\theta}) \, p_\mathrm{clustering}(\bm{\theta}) \, \mathrm{d}\bm{\theta}$. As the problem is two-dimensional, we estimate posterior densities from samples using kernel density estimation. We apply a Jacobian transformation to reweight the posterior samples of~\citetalias{Stiskalek_2021} (which assumed a flat prior on $\alpha$) to a common flat prior on $\tan^{-1}\alpha$.

\subsection{Generating mock data}

We draw independent mock datasets from the best-fitting parameter vector $\hat{\bm{\theta}}$ by re-running the full forward model and re-drawing all stochastic model elements (SHAM scatter, stellar/gas masses, halo-galaxy matching, and rotation curve computation). Because the selection step can reject a galaxy--halo pair in a given realisation, some galaxies may not be successfully matched to a halo. We handle this via rejection sampling: for each galaxy, we repeat the forward draw with independent random seeds until it is accepted by the selection, and retain the first accepted $\Vmax$. This yields one $\Vmax$ per galaxy sampled from the model's conditional predictive distribution conditional on selection.

To emulate measurement noise, we perturb each simulated $\Vmax$ using the fractional SPARC uncertainties. For the $i$\textsuperscript{th} galaxy, we set
\begin{equation} \label{eq:obs_noise_1}
    \sigma_{V_{\max},i} \;=\; V_{\max,i} \frac{\sigma_{V_{\max},i}^{\text{(SPARC)}}}{V_{\max,i}^{\text{(SPARC)}}}\, ,
\end{equation}
and draw
\begin{equation} \label{eq:obs_noise_2}
    V_{\max,i}^{\star} \;\sim\; \mathcal{N}\bigl(V_{\max,i},\;\sigma^2_{V_{\max},i}\bigr).
\end{equation}


\section{Results}\label{sec:results}

We present posterior constraints from a baseline model without halo selection (\cref{sec:results_nosel}), then introduce selection on $\Vmaxhalo$ and compare model evidences (\cref{sec:results_sel}). Mock tests validating the inference pipeline are presented in \cref{sec:mocks}.

\subsection{Model without selection}\label{sec:results_nosel}

\begin{figure*}
    \centering
    \includegraphics[width=\textwidth]{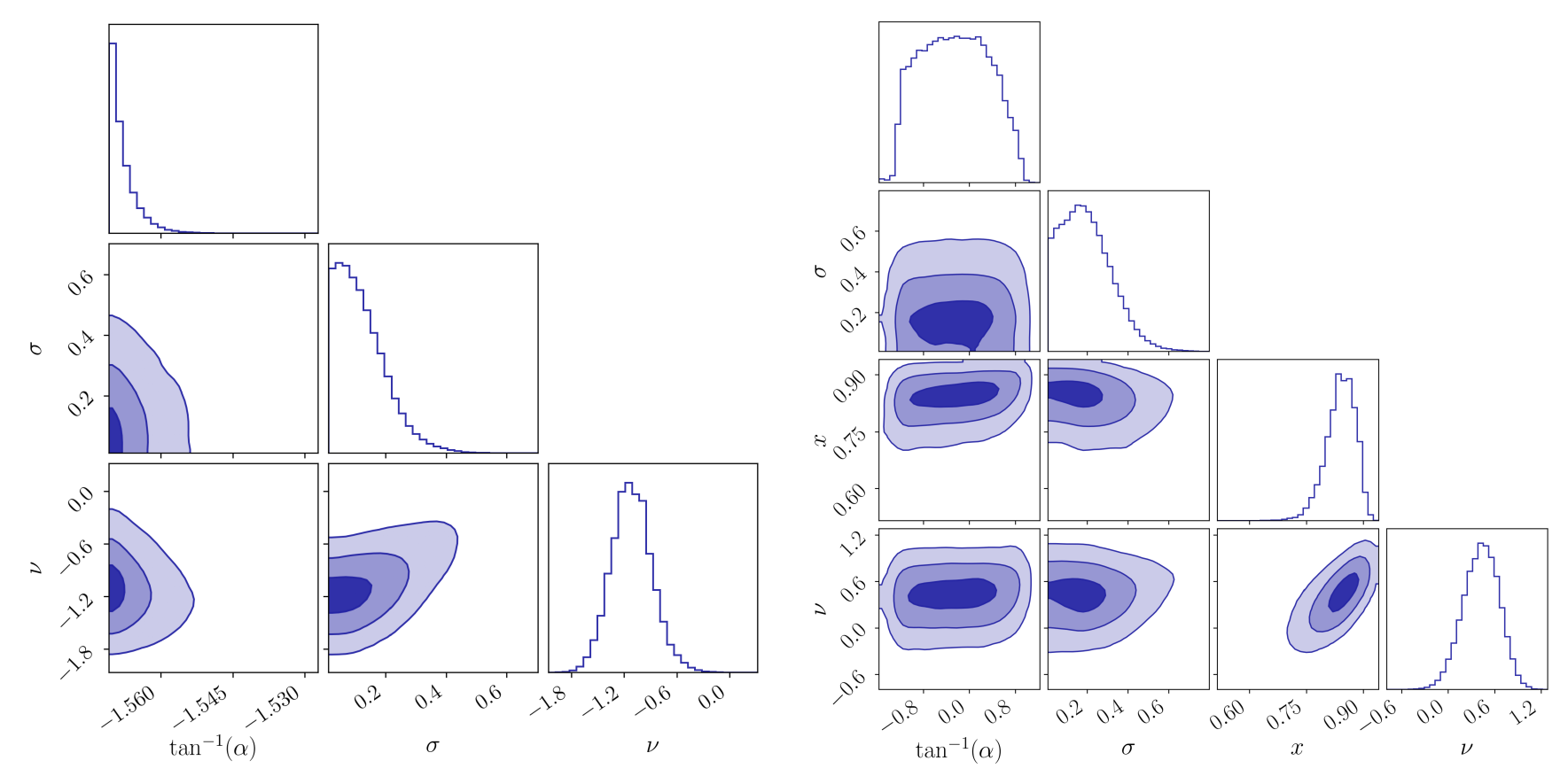}
    \caption{Posterior constraints on the three-parameter model (\emph{left}) and the four-parameter model with selection on halo properties (\emph{right}). Contours enclose 39.3, 86.5, and 98.9 per cent of posterior mass (corresponding to $1\sigma$, $2\sigma$, and $3\sigma$ for a 2D Gaussian). Without selection, the data favour extreme SHAM parameters in tension with independent constraints from galaxy clustering. Introducing selection brings all parameters into agreement with clustering, but requires a strong selection threshold ($x = 0.84 \pm 0.04$), meaning SPARC-like galaxies occupy the lowest ${\approx}16$ per cent of the $\Vmaxhalo$ distribution at fixed $\Mvir$.}
    \label{fig:corner_plots}
\end{figure*}

The left panel of Figure~\ref{fig:corner_plots} shows the posterior constraints for the three-parameter model in which SPARC galaxies are assumed to be an unbiased draw from the full halo population ($\bm{\theta} = \{\tan^{-1}\alpha, \,\sint, \,\nu\}$). Three features emerge:
\begin{itemize}[align=parleft, left=0pt, labelsep=0.5em]
    \item Matching proxy: The posterior for $\tan^{-1}\alpha$ is prior-bound at the lower limit ($\tan^{-1}\alpha < -1.56$ at $2\sigma$), implying $\alpha < -92.6$, i.e.\ a matching proxy that heavily suppresses satellite haloes.
    \item Scatter: The posterior peaks at $\sint = 0$ with a $1\sigma$ upper bound of ${\sim}0.2$ dex, lower but consistent with the ${\sim}0.2$ dex typically inferred for massive galaxies from clustering analyses~\citep{More_2010,Reddick_2013,Lehmann_2017}. (An extension to mass-dependent scatter is discussed in Section~\ref{sec:discussion_interpretation}.)
    %
    \item Halo response: The data favour expanded haloes, $\nu = -1.12^{+0.23}_{-0.22}$, corresponding to a reversal of the standard adiabatic contraction of comparable magnitude.
\end{itemize}
These values are in tension with independent clustering constraints, which favour $\alpha \gtrsim 0$ and $\sint \approx 0.2$ dex for massive galaxies~\citepalias{Stiskalek_2021}. SPARC targets \HI-rich, late-type galaxies, which may not be representative of the full halo population at fixed stellar mass. To account for this, we introduce a halo-level selection on $\Vmaxhalo$ at fixed $\Mvir$ (Section~\ref{sec:selection}) and repeat the inference.

\subsection{Model with selection}\label{sec:results_sel}

\begin{figure*}
    \centering
    \includegraphics[width=0.9\textwidth]{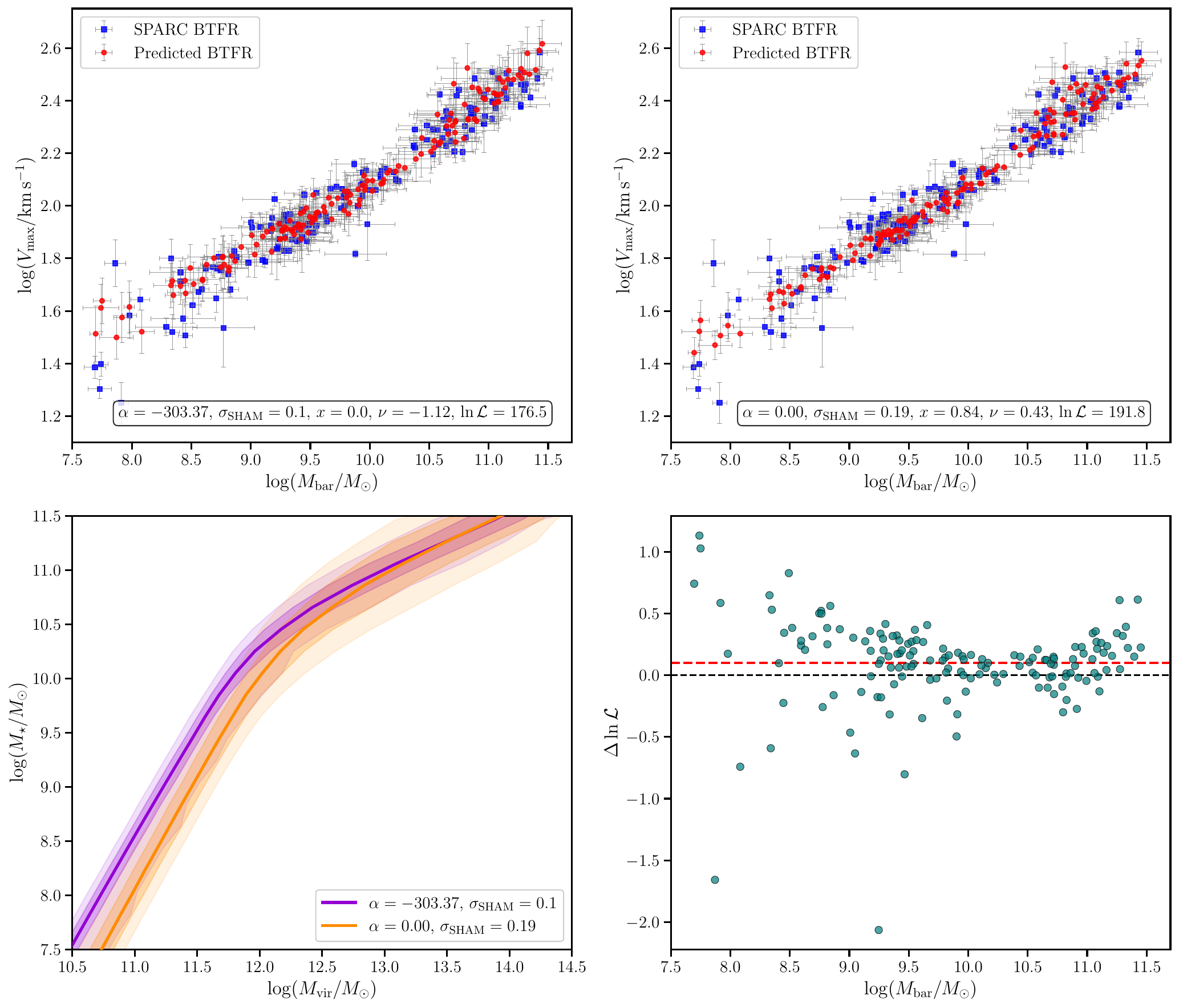}
    \caption{Comparison of the two best-fit models: the baseline 3-parameter model and the 4-parameter model with selection. \textbf{Top row:} BTFRs predicted by the two models overlaid on the observed SPARC data. \textbf{Bottom left:} The inferred stellar-to-halo mass relation for both models, showing the median, $1\sigma$ and $2\sigma$ credible intervals. \textbf{Bottom right:} The difference in log-likelihood per galaxy ($\Delta \ln \mathcal{L} = \ln \mathcal{L}_{\mathrm{sel}} - \ln \mathcal{L}_{\mathrm{base}}$) as a function of baryonic mass. Points above the dashed reference line favour the selection model. The red dashed line indicates the mean log-likelihood difference, $\ln \mathcal{L}_{\text{mean}} = 0.1$.}
    \label{fig:btfr_panel}
\end{figure*}

The right panel of Figure~\ref{fig:corner_plots} shows the four-parameter posterior:
\begin{itemize}
    \item Matching proxy: The constraint on $\alpha$ weakens substantially and shifts toward zero ($-0.8979 < \tan^{-1}\alpha < 0.3826$ at $68$ per cent confidence), removing the tension with clustering constraints.
    \item Scatter: The preferred scatter increases to $\sint = 0.19^{+0.13}_{-0.11}$ dex, consistent with the ${\sim}0.2$ dex for massive galaxies from clustering analyses (though also with zero scatter).
    \item Halo response: The posterior shifts to $\nu = 0.43^{+0.22}_{-0.24}$, indicating mild net contraction or no significant response.
    \item Selection: The selection parameter is tightly constrained to $x = 0.84 \pm 0.04$, meaning SPARC-like galaxies are drawn from the lowest ${\approx}16$ per cent of the $\Vmaxhalo$ distribution at fixed $\Mvir$.
\end{itemize}
Introducing selection thus removes the tension in $\alpha$ between our work and clustering constraints, and shifts $\nu$ to values consistent with hydrodynamical simulations, but requires a very strong $\Vmaxhalo$-based selection.
Figure~\ref{fig:btfr_panel} compares the predicted BTFR from the baseline and selection models to the SPARC data: visually, both models successfully reproduce the tight correlation, slope, and scatter of the observed BTFR, but the selection model achieves this with parameters consistent with independent clustering constraints. (There may be a hint of curvature in the predicted relation in both cases, as was found to be a problem for the SPARC BTFR in~\citealt{Desmond2017}.) The bottom left panel shows the distinct stellar-to-halo mass relations implied by the two models. While they converge at the high-mass end, they diverge at lower masses. The specific shape of the baseline relation is warped by the extreme $\alpha$ proxy required to match the observed BTFR in the absence of explicit selection. The bottom right panel shows the difference in log-likelihood $\Delta \ln\mathcal{L}$ between the two models for each individual galaxy as a function of baryonic mass. The selection model generally improves the fit at both the low-mass ($M_\text{bar} < 10^9 M_\odot$) and high-mass ($M_\text{bar} > 10^{11} M_\odot$) ends of the sample, likely due to decreased variance in the simulated $\Vmax$ distributions from the selection threshold. By truncating the halo population, the selection model eliminates the high-velocity tail (see Figure~\ref{fig:velocity_distribution_example}), narrowing the spread of predicted velocities for a given stellar mass and concentrating probability mass more tightly around the observed data.

Using the \texttt{harmonic} code~\citep{polanskaLearnedHarmonicMean2024}, we find $\Delta \ln \mathcal{Z} = 15.66$ and a Bayes factor of $6\times10^{6}$ in favour of the model that includes selection, indicating a decisive preference for including this additional parameter.

\subsection{Comparison to clustering constraints}

\begin{figure*}
    \centering
    \includegraphics[width=\textwidth]{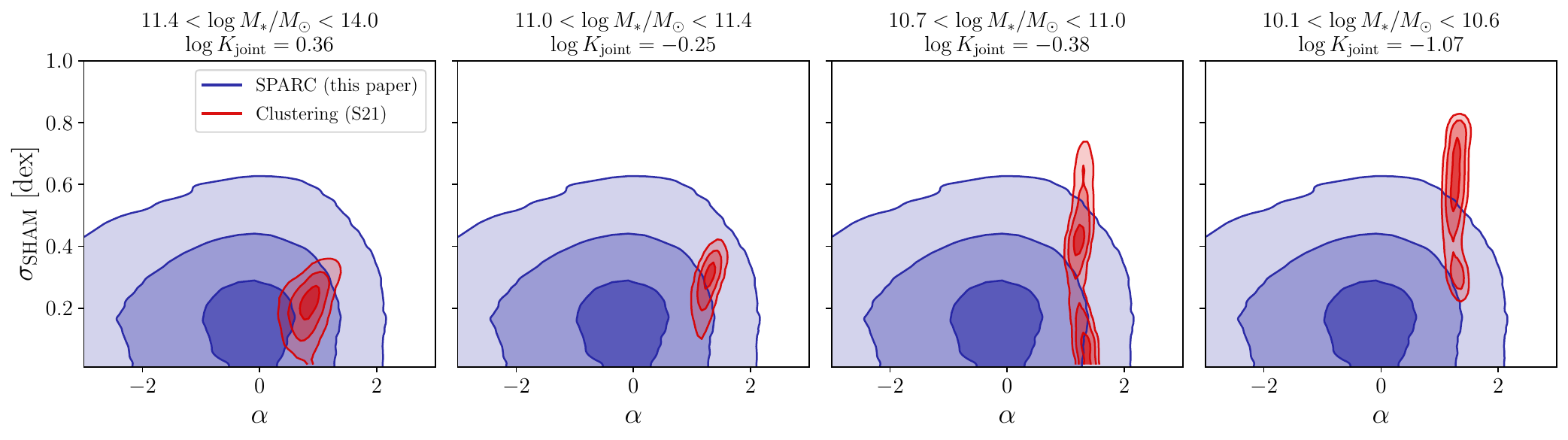}
    \caption{A comparison of the posterior constraints on the SHAM parameters from our SPARC analysis with selection (which assumes mass-independent SHAM parameters), with independent constraints from the clustering analysis of~\citetalias{Stiskalek_2021}, for their four stellar mass bins (shown above the panels). All posteriors are reweighted to a common flat prior on $\tan^{-1}\alpha$. Contours enclose 39.3, 86.5, and 98.9 per cent of posterior mass (corresponding to $1\sigma$, $2\sigma$, and $3\sigma$ for a 2D Gaussian). The Bayes factor comparing the joint model (in which both datasets constrain the same underlying parameters) to independent models is shown above each panel, with the highest three stellar mass bins showing no more than weak evidence for tension, but the lowest mass bin (which is closest to the SPARC mean mass) indicating strong tension.}
    \label{fig:compare_posteriors_alpha}
\end{figure*}

Figure~\ref{fig:compare_posteriors_alpha} compares our posterior constraints on $(\alpha, \sint)$ from the model including selection with those from the independent clustering analysis of~\citetalias{Stiskalek_2021}, which constrains the same SHAM parameters independently in four stellar mass bins. Specifically, we compare to their \emph{optically-selected} sample constraints: our model assumes universal SHAM parameters for all galaxies, with \HI-rich systems then selected based on halo properties, rather than treating \HI-selected galaxies as a distinct population with their own AM parameters. We use their optically-selected sample with NSA Petrosian stellar masses for consistency with our stellar masses. We compute Bayes factors $K_\mathrm{joint}$ (Equation~\ref{eq:Kjoint}) to quantify the consistency between the two datasets. For the three highest mass bins ($\log(M_\star/\mathrm{M}_\odot) > 10.7$), we find $\log K_\mathrm{joint} = +0.36$, $-0.25$, and $-0.38$ respectively---all inconclusive or showing weak evidence for consistency on the Jeffreys scale. This indicates quantitatively that once halo selection is included, clustering and kinematics require consistent abundance-matching galaxy--halo connections at these masses. However, the lowest mass bin ($\log(M_\star/\mathrm{M}_\odot) = 10.1$--$10.6$), which overlaps most closely with the SPARC sample, yields $\log K_\mathrm{joint} = -1.07$, showing strong tension.

\section{Discussion}\label{sec:discussion}

We interpret our results in the context of independent constraints and physical expectations (\cref{sec:discussion_interpretation}), discuss caveats and systematic uncertainties (\cref{sec:discussion_caveats}), and outline directions for future work (\cref{sec:discussion_future}).

\subsection{Interpretation and broader ramifications}\label{sec:discussion_interpretation}

A key innovation in our work is the development and use of a Bayesian forward model for constraining the galaxy--halo connection using kinematics that operates on a galaxy-by-galaxy level. By forward-modelling the data directly we can account for relevant effects in the direction in which they operate physically, and by making predictions for individual galaxies we ensure that the maximum amount of information in the data is utilised in the inference. This contrasts with methods that first summarise the data into scaling relations such as the TFR, which causes information loss.

The baseline three-parameter model (Section~\ref{sec:results}) yields extreme parameter values that, while reproducing the dynamics reasonably well, are difficult to reconcile with independent constraints. The inferred halo expansion ($\nu \approx -1$) reverses the standard adiabatic contraction by a comparable factor, consistent with previous AM-based analyses of the (stellar mass) TFR~\citep{Desmond_2015}, which found that reproducing the observed normalisation requires weak contraction or net expansion. The SHAM scatter posterior peaks near zero, driven by the tight observed BTFR ($\sim$0.1 dex intrinsic scatter), though the constraint is broad and remains consistent at the $2\sigma$ level with the $\sim$0.2 dex typically inferred for massive galaxies from galaxy clustering and satellite fractions
\citep{Reddick_2013,Stiskalek_2021}. Most strikingly, the matching proxy parameter $\alpha$ is pinned to its prior bound, implying a prescription that heavily suppresses satellites at fixed stellar mass.

Physically, reduced star formation efficiency in satellite haloes is plausible: environmental processes such as ram-pressure stripping and strangulation can quench satellites, making them under-luminous relative to centrals of the same halo mass~\citep{Peng_2010, Wetzel_2013}. However, the extreme $\alpha$ values preferred by the baseline model are in strong tension with clustering constraints, which favour $\alpha \gtrsim 0$~\citepalias{Stiskalek_2021}. This tension suggests that the baseline model may be missing an important ingredient.

SPARC is not a representative optically-selected, stellar-mass-limited sample: it targets \HI-rich, late-type galaxies that may preferentially reside in less dense environments. Without accounting for selection effects, the model may absorb such population differences into $\alpha$ and $\nu$, biasing both toward extreme values. In this interpretation, the baseline model compensates for unmodelled selection by (i) expanding haloes to lower $\Vmax$ at fixed baryonic mass, (ii) minimising SHAM scatter, and (iii) suppressing satellites that form stars efficiently.
Introducing halo-level selection---expected a priori to some degree---largely resolves these tensions (Section~\ref{sec:results}). The matching proxy becomes consistent with $\alpha \gtrsim 0$, and the halo response shifts from strong expansion to mild contraction or no net response.

In~\Cref{fig:compare_posteriors_alpha} we compared our posterior for the model with selection to those of~\citetalias{Stiskalek_2021} for the optically-selected sample. That analysis constrained AM parameters for \HI-selected galaxies as we do here, using an \HI-specific stellar mass function and treating \HI-rich systems as a distinct population with their own SHAM parameters. Our approach differs conceptually: we assume all galaxies are populated according to a universal SHAM prescription, with survey-specific samples then arising from selection on halo properties. This avoids having to specify which fraction of the total simulated haloes appear in the \HI{} survey before the abundance matching step, which is computationally expensive as it requires repeating abundance matching for each separate value of the selection parameter. Under this framework, the appropriate comparison is to optically-selected constraints, which should reflect the underlying universal relation we assume. 

The best-fit selection parameter $x \approx 0.84$ indicates that SPARC-like galaxies occupy the lowest ${\sim}16$ per cent of the $\Vmaxhalo$ distribution at fixed virial mass. Because $\Vmaxhalo$ at fixed $\Mvir$ correlates with concentration and formation time, this suggests that SPARC preferentially samples low-concentration, late-forming haloes. The general direction of this preference is physically plausible: gas-rich, actively star-forming disc galaxies are expected to reside in haloes that formed more recently and have shallower central potentials, allowing them to retain their cold gas reservoirs. This conclusion echoes~\citet{Desmond_2015}, who argued that selection effects preferentially eliminate fast-rotating galaxies from kinematic samples, requiring late-type spirals to inhabit low-concentration haloes (around the lowest ${\sim}50$ per cent) to match the observed BTFR normalisation. While they modelled this selection through the correlation between disc inclination and line-of-sight velocity, we parametrise it directly at the halo level; both approaches converge on the same physical picture. 

Various studies have found differences in the properties of the haloes occupied by early and late-type galaxies. \citet{wojtakPhysicalPropertiesUnderlying2012} and \citet{mandelbaumStrongBimodalityHost2016} find blue galaxies reside in lower concentration haloes at fixed halo mass, and the \texttt{UniverseMachine} model \citet{Behroozi2019} finds star-forming galaxies tend to live in haloes that are still accreting mass now. Both of these are qualitatively consistent with our $\Vmax$ selection. However, these and other studies \citep{rodriguez-pueblaStellartohaloMassRelations2015,mandelbaumStrongBimodalityHost2016} have found blue and red galaxies occupy different SHMRs, whereas we assume a universal SHAM relation. In the future our framework should be extended to allow different SHAM relations by galaxy-type, in addition to selection.

Both our study and~\citet{Desmond_2015} require strong selection, with our tighter threshold (${\sim}16$ vs ${\sim}50$ per cent) likely reflecting a combination of the smaller, more targeted sample and more constraining per-galaxy likelihood framework employed here. Whether such a tight connection between halo properties and \HI\ content exists in nature remains unknown. The $M_\mathrm{HI}$--$M_\star$ relation exhibits substantial scatter~\citep[$\sim$0.3--0.4 dex;][]{Pan2021, Pan2025}, suggesting that gas content is not tightly determined by stellar mass alone. It would therefore be surprising if $\Vmaxhalo$ at fixed $\Mvir$---or any other proxy of the halo mass distribution---could strongly predict \HI\ content. One possibility is that the extreme selection parameter reflects a weaker underlying correlation: if $\Vmaxhalo$ is only loosely related to the observables that actually drive SPARC selection, a strong threshold on $\Vmaxhalo$ may be needed to approximate a more moderate selection on \HI\ mass. Other possible candidates include whether a galaxy is a central or a satellite and its merger history (both of which may predict the disruption of either the disc structure and/or the gas reservoir). Future analyses of samples with well-characterised selection functions in terms of \HI\ observables will be essential to assess whether halo-level selection can reconcile kinematic and clustering constraints in a physically realistic way.

An important issue is the potential mass-dependence of SHAM scatter. Our model assumes a single, mass-independent $\sint$, but the clustering analyses of~\citetalias{Stiskalek_2021} find that scatter varies significantly with stellar mass: they infer $\sigma_\mathrm{AM} \approx 0.61^{+0.11}_{-0.14}$ dex for $10.1 < \log(M_\star/M_\odot) < 10.6$, decreasing to $\sigma_\mathrm{AM} \approx 0.22^{+0.05}_{-0.05}$ dex for $11.4 < \log(M_\star/M_\odot) < 14.0$ for optically-selected samples. This creates a potential tension: SPARC predominantly samples lower-mass galaxies, where clustering predicts the largest SHAM scatter, yet the observed BTFR remains tight ($\sim$0.1 dex intrinsic scatter) across the full mass range. If the galaxy--halo connection truly has $\sim$0.5 dex scatter at low masses, it is unclear how SPARC kinematics can be so well-behaved---selection alone cannot easily suppress scatter that is intrinsic to the SHAM relation. A binned analysis of SPARC galaxies in stellar mass could help disentangle these effects, though it would further complicate the selection modelling. Extending the framework to mass-dependent $\sint(M_\star)$ will be necessary to fully assess whether kinematics and clustering can be jointly satisfied.

Relatedly, \citet{Maccio2020} tested AM predictions using dynamical masses for 190 Virgo cluster galaxies spanning $10^8 < M_\star/\mathrm{M}_\odot < 10^{11}$, finding that observed scatter exceeds AM predictions by a factor of ${\sim}5$; however, they argued this was dominated by measurement uncertainties, illustrating the difficulty of finding well-controlled kinematic samples for such tests. Other clustering-based studies have found no evidence for mass-dependent scatter in the galaxy--halo connection~\citep[e.g.][for haloes down to $10^{11.5}\,\mathrm{M}_\odot$]{Mitra2025}.
Another possibility is a framework such as Modified Newtonian Dynamics (MOND;~\citealt{Milgrom_1}) which postulates a direct connection between baryons and dynamics and hence implies negligible scatter in the BTFR (for reviews see~\citealt{Famaey_McGaugh,Banik_Zhao,Famaey_Durakovic,Desmond_MOND}).

Hydrodynamical simulations such as EAGLE~\citep{Schaller2015}, Illustris/TNG~\citep{Genel2014, Pillepich2018}, and FIRE~\citep{Hopkins2018} offer an alternative to the simple semi-empirical prescriptions employed here, producing direct predictions for galaxy dynamics. However, these simulations are calibrated on ensemble statistics (stellar mass functions, sizes, star formation rates) rather than individual galaxy kinematics. Recent work attempting to constrain hydrodynamical simulation parameters with observations has similarly relied on population-level properties such as the dark matter fraction as a function of stellar mass~\citep{Busillo2023, Busillo2025}, rather than per-galaxy velocity measurements. By employing simple semi-empirical constructions, we match the observed stellar and gas properties used as inputs to the forward model, allowing us to isolate which additional ingredients---such as halo response driven by baryonic feedback~\citep[e.g.][]{Pontzen_Governato_2012} and sample selection---are required to reproduce the kinematic data, without relying on specific recipes for star formation and feedback. Hydrodynamical simulations also remain limited in volume and resolution, making it challenging to generate large mock samples that match the observed stellar mass function while resolving internal kinematics~\citep{Crain2023}. Our semi-empirical approach therefore provides a complementary method to extract information from galaxy kinematics and constrain the galaxy--halo connection, in which phenomenological ingredients can be investigated individually.

\subsection{Caveats and systematics}\label{sec:discussion_caveats}

Several simplifying assumptions underlie our model. First, we adopt a mass-independent SHAM prescription: the proxy parameter $\alpha$ and scatter $\sint$ are constant across the stellar mass range probed by SPARC. As discussed in Section~\ref{sec:discussion_interpretation}, clustering analyses find evidence for mass-dependent scatter, increasing toward lower masses~\citepalias{Stiskalek_2021}. If such trends are present, our single-valued parameters represent effective averages that may not accurately describe the edges of the mass distribution.

Second, our halo-level selection model is deliberately simple. The threshold parameter $x$ imposes a sharp cut on $\Vmaxhalo$ at fixed $\Mvir$, effectively assuming a tight (deterministic) mapping between halo properties and \HI\ content. In reality, the connection between halo concentration, formation time, and cold gas mass is likely to have substantial scatter, modulated by environment, merger history, and stochastic feedback~\citep{Stiskalek2022scatter}. A more realistic treatment would introduce a probabilistic selection function with its own free parameters, at the cost of additional model complexity.

We assume an NFW density profile for all haloes, modified only by the global response parameter $\nu$. A single-parameter halo response may be a strong simplification: hydrodynamical simulation studies have found the degree of contraction or expansion varies with radius, stellar-to-halo mass ratio, and assembly history in ways that a global $\nu$ cannot fully capture~\citep{DiCintio2014,Tollet2016,Velmani2022}. We also assume spherical symmetry, whereas $N$-body simulations generically produce triaxial haloes; depending on the alignment of discs, triaxial haloes may lead to $\Vmax$ variations of order ${\sim}10$ per cent~\citep{HayashiNavarro2006}. These effects could introduce additional scatter or systematic shifts in the predicted rotation velocities.

We adopt the SPARC mass-to-light model, which assumes a fixed mass-to-light ratio in the $3.6~\mu$m band for the bulge and disc. Spatially varying mass-to-light ratios---arising from radial gradients in stellar age, metallicity, or dust---could alter both the inferred total stellar mass and the shape of the stellar contribution to the rotation curve. Recent work using full stellar population synthesis modelling suggests that such variations do occur and can shift the inferred dark matter content~\citep{Varasteanu2025}. Fits to the Radial Acceleration Relation also prefer somewhat different mass-to-light ratios~\citep{Desmond_Cassini}. Systematic differences in $M_\star$ would also propagate into our SHAM assignment and halo response inferences.

Finally, our dynamical model treats both the stellar and gas components as infinitesimally thin exponential discs with identical scale lengths. Real discs have finite vertical extent, which slightly reduces the mid-plane circular velocity at fixed mass and size. We also neglect pressure support: for gas-rich, low-mass systems the observed rotation velocity can underestimate the true circular velocity due to asymmetric drift, and our model does not apply any such correction. Both effects are likely subdominant for the SPARC sample but could become more important for lower-mass or more dispersion-supported systems. For high-mass systems, a bigger concern is the simple treatment of bulges as spherical components; real bulges are often flattened or triaxial, which would alter their contribution to the rotation curve~\citep{Fisher2008,Costantin2018,Desmond_Cassini}.

\subsection{Future work}\label{sec:discussion_future}

A natural next step is to apply this framework to samples with well-defined selection criteria. The MIGHTEE-\HI\ survey~\citep{Maddox2021,Ponomareva2021} provides homogeneous photometry and kinematics for \HI-selected galaxies, including linewidth measurements that serve as a proxy for $\Vmax$ whilst also extending such analyses to $z \sim 0.4$~\citep{Jarvis2025}. Applying our forward model to such samples would reduce systematic uncertainties associated with selection modelling and enable more robust inferences on the galaxy--halo connection. Wider field surveys such as ALFALFA~\citep{Haynes2018} could also be used, but at the expense of less precise photometry and auxiliary parameters such as inclination.

The selection model itself could be made more flexible. Rather than a sharp threshold on a single halo property, one could adopt probabilistic selection functions with scatter, or selection on multiple halo properties such as concentration or environment. Empirical relations between \HI\ mass and stellar mass~\citep{Pan2021} could inform such selection functions and help connect galaxy observables to halo properties.

Future \HI\ surveys will dramatically expand the available samples. WALLABY on ASKAP~\citep{Koribalski2020}, MIGHTEE-\HI\ on MeerKAT, and eventually the SKA~\citep{Blyth2015} will provide \HI\ detections and linewidths for millions of galaxies extending to higher redshift. Applying our framework to these samples would enable evolutionary studies of the galaxy--halo connection and baryonic effects on halo structure across cosmic time.

The halo response model could be improved by adopting more physically-motivated prescriptions from hydrodynamical simulations. \citet{DiCintio2014} and \citet{Tollet2016}, for example, parameterise halo response as a function of stellar-to-halo mass ratio and predict both contraction and expansion depending on galaxy properties. More recent analytic models~\citep{liResponseDarkMatter2022} trace energy diffusion during gas ejection events and could provide a more physical basis for the response parameter. Incorporating such models would allow the halo response to depend on additional galaxy properties such as star-formation rate or gas fraction, rather than being a single global parameter.

Resolved rotation curve data contain yet more information---the Radial Acceleration Relation~\citep[e.g.][]{McGaugh2016,Desmond2017,Stiskalek2023RAR,desmondUnderlyingRadialAcceleration2023,Desmond_Cassini,Varasteanu2025} demonstrates a tight correlation between observed and baryonic accelerations at each radius---and extending our framework to exploit this is a natural direction for future work. Another extension of this work is to the Milky Way satellite population, where forward-modelling pipelines have been used to constrain the galaxy--halo connection and satellite disruption from photometric census data~\citep{Nadler2019,Nadler2020}, and semi-analytic frameworks provide physically-motivated ways to predict satellite properties from halo assembly histories~\citep{Kravtsov2022,Manwadkar2022}. Kinematics has begun to enter these analyses in compressed form, such as completeness-corrected velocity dispersion distributions, enabling constraints on small-scale dark matter physics~\citep{Kim2018,Kim2021,Esteban2023}. Our per-galaxy likelihood approach suggests a route to sharpen these constraints by forward-modelling the full line-of-sight velocity distributions of individual satellites within a similar Bayesian framework. This would leverage the full information content of stellar kinematics to probe dark matter structure and galaxy formation in the smallest haloes, and enable direct comparison to our study of predominantly isolated, late-type galaxies.

Finally, the likelihood framework could be adapted to constrain any model that predicts galaxy observables---such as hydrodynamical simulations or semi-empirical models like \textsc{UniverseMachine}~\citep{Behroozi2019}---maximising the constraining power of these data compared to population-level summary statistics.

\section{Conclusion}\label{sec:conclusion}

We present a Bayesian forward model for the maximum rotation velocities of SPARC galaxies, combining subhalo abundance matching with a parameterised halo response and constraining model parameters. Our likelihood is defined per galaxy, maximising constraining power by preventing information loss through compression of the data into summary statistics (e.g. the slope and intercept of the BTFR). Our conclusions are as follows:
\begin{itemize}
    \item A baseline model without sample selection reproduces the observed $\Vmax$ distribution but requires extreme parameter values: strong halo expansion ($\nu = -1.12^{+0.23}_{-0.22}$), low SHAM scatter ($\sint < 0.15$ dex at $1\sigma$; a consequence of the BTFR's low intrinsic scatter), and a SHAM proxy pinned to its lower prior bound ($\alpha < -92.6$), corresponding to the maximal suppression of stellar mass in stripped haloes possible within our SHAM parameterisation. This is in severe tension with independent clustering constraints (Figure~\ref{fig:corner_plots}), which favour $\alpha > 0.0$.
    \item Introducing halo-level selection on $\Vmaxhalo$ at fixed $\Mvir$ is decisively favoured (Bayes factor $6\times10^{6}$) and brings SHAM parameters into agreement with clustering: $\sint = 0.19^{+0.13}_{-0.11}$ dex and $\nu = 0.43^{+0.22}_{-0.24}$. The selection threshold $x = 0.84 \pm 0.04$ implies SPARC galaxies occupy the lowest ${\sim}16$ per cent of $\Vmaxhalo$ at fixed $\Mvir$ (which correlates with preferential sampling of low-concentration, late-forming haloes).
    \item The inferred selection is extremely strong: given the large scatter in the $M_\mathrm{HI}$--$M_\star$ relation, it is unclear whether $\Vmaxhalo$ can predict gas content precisely enough; alternatively, $x$ may absorb model misspecification from unmodelled baryonic physics or non-standard dark matter phenomenology.
    \item Independent clustering constraints imply $\sigma_\mathrm{AM} \sim 0.5$ dex at low stellar masses~\citepalias{Stiskalek_2021}, yet the BTFR remains tight across this range, suggesting residual tension. Extending the framework to mass-dependent scatter will be necessary to fully test whether kinematics and clustering can be convincingly reconciled across the mass range in which they overlap.
\end{itemize}

Our analysis demonstrates the constraining power of resolved rotation curves on the galaxy--halo connection, and the importance of modelling sample selection when combining kinematic and clustering constraints. Forward-modelling frameworks of this type will be crucial for unlocking the full constraining power of upcoming \HI\ surveys such as MIGHTEE, WALLABY and the SKA, which are set to provide linewidths for millions of galaxies to high redshift in the coming years. This will constrain halo structure, provide information on dark matter microphysics, test empirical models of the galaxy--halo connection and trace the co-evolution of baryonic and dark matter over cosmic time.




\begin{figure*}
    \centering
    \includegraphics[width=0.9\textwidth]{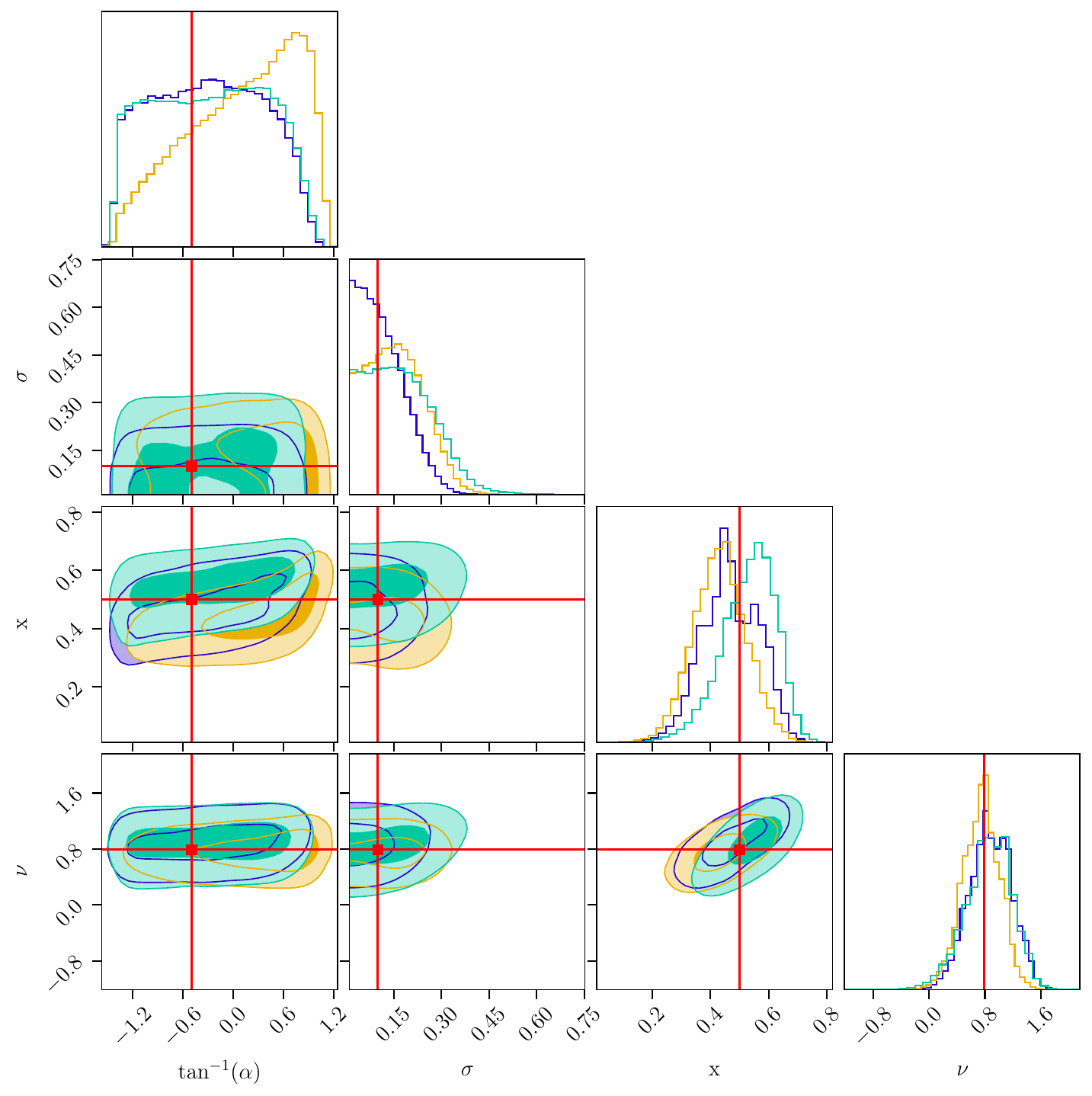}
    \caption{Mock validation of the inference pipeline. Contours enclose 39.3 and 86.5 per cent of the 2D posterior ($1\sigma$ and $2\sigma$ for a 2D Gaussian). We show constraints for three independent mock datasets generated at $\bm{\theta}_\mathrm{true} = (\tan^{-1}\alpha, \sint, x, \nu) = (-0.5, 0.1, 0.5, 0.789)$. The red lines indicate these truths. The true parameters fall within the $1\sigma$ contours in most panels, demonstrating that the pipeline produces well-calibrated posteriors.}
    \label{fig:mock_test_corner}
\end{figure*}

\section*{Data availability}

Data from the SPARC galaxy survey are publicly available at
\url{http://astroweb.cwru.edu/SPARC/},
the \texttt{Uchuu} suite of $N$-body simulations at
\url{https://skiesanduniverses.org/Simulations/Uchuu/},
and the NSA catalogue at
\url{https://live-sdss4org-dr13.pantheonsite.io/}.
All other data and computer code underlying this study will be made available upon reasonable request to the corresponding author.

\appendix

\section{Mock tests}\label{sec:mocks}

We validate the inference pipeline using 50 mock datasets generated from fiducial parameters ($\tan^{-1}\alpha{=} -0.5$, $\sint {=} 0.1$, $x{=} 0.5$, $\nu{=}0.789$) with observational noise added as per Equations~\eqref{eq:obs_noise_1} and~\eqref{eq:obs_noise_2}. Each mock is analysed with the same priors and pipeline as the real data. Figure~\ref{fig:mock_test_corner} shows posterior constraints for three realisations: the true parameter values (solid lines) lie within the $1\sigma$ credible regions in most cases. Across all 50 mocks, the distribution of posterior residuals relative to the injected values is consistent with the expected coverage, confirming that the pipeline is well-calibrated.

Comparing real and mock posteriors, the constraint on $\alpha$ is of similar width in both, $\sint$ is slightly more constrained in the mocks and $x$ is much more constrained in the real data than in the mocks. This discrepancy suggests there may be additional sources of systematic and/or statistical uncertainty present in the real data compared to the mocks.

\section*{Acknowledgements}

We thank Anastasia Ponomareva, Andreea Varasteanu and Federico Lelli for useful inputs and discussion.

FB, TY and MJJ acknowledge support from UKRI Frontiers Research Grant [EP/X026639/1], which was selected by the ERC.
FB also acknowledges support from the Oxford University Astrophysics Summer Research Programme.
HD is supported by a Royal Society University Research Fellowship (grant no. 211046).
RS acknowledges financial support from STFC Grant No. ST/X508664/1 and the Snell Exhibition of Balliol College, Oxford.

We thank Jonathan Patterson for smoothly running the Glamdring Cluster hosted by the University of Oxford, where the data processing was performed.

\bibliographystyle{mnras}
\bibliography{main} 

\bsp
\label{lastpage}
\end{document}